\documentclass[11pt,a4paper]{article}
\usepackage{mathptmx}
\usepackage{fullpage}
\usepackage{microtype}
\usepackage{rotating}
\usepackage[utf8]{inputenc}
\usepackage{amssymb}
\usepackage{amsmath}
\usepackage{graphicx}
\usepackage{url}
\usepackage{multirow}
\usepackage{amsthm}
\usepackage[ruled,vlined,linesnumbered]{algorithm2e}

\newcommand{\real}{\mathbb{R}}
\newcommand{\nat}{\mathbb{N}}
\newcommand{\Oh}{\mathrm{O}}

\newcommand{\SP}{\mathrm{SP}}

\renewcommand{\subparagraph}[1]{\paragraph{\normalfont\itshape{#1}}}

\newtheorem{mytheorem}{Theorem}
\newtheorem{mycorollary}[mytheorem]{Corollary}

\title{Compressed Transmission of Route Descriptions%
\thanks{Partially supported by DFG grant SA 933/5-1}%
}

\author{%
  Gernot Veit Batz, Robert Geisberger, Dennis Luxen, and Peter Sanders\\\normalsize
  {\tt\{batz,geisberger,luxen,sanders\}@kit.edu}
}

\SetKwBlock{KwFunction}{function}{end function}
\newcommand{\Function}[4]{\KwFunction(\mbox{\textit{#1}$($#2$)$ : #3}){#4}}

\SetKwBlock{KwProcedure}{procedure}{end procedure}

\begin{document}

\maketitle

\begin{abstract}
\noindent
We present two methods to compress the description of a route in a road network, i.e., of a path in a directed graph.
The first method represents a path by a sequence of \emph{via edges\nocorr}.
The subpaths between the via edges have to be unique shortest paths.
Instead of via edges also \emph{via nodes} can be used, though this requires some simple preprocessing.
The second method uses contraction hierarchies to replace subpaths of the original path by shortcuts.
The two methods can be combined with each other.
Also, we propose the application to mobile server based routing:
We compute the route on a server which has access to the latest information about congestions for example.
Then we transmit the computed route to the car using some mobile radio communication.
There, we apply the compression to save costs and transmission time.
If the compression works well, we can transmit routes even when the bandwidth is low.
Although we have not evaluated our ideas with realistic data yet, they are quite promising.

\end{abstract}


\section{Introduction}

Today GPS-based car navigation is quite common.
There, routes can be computed either by a mobile device located in the car  or by a server system located in a computing center.
The latter can be useful to take temporary congestions or the latest changes of the road network or of the POI data into account.
But then, a description of the route has to be transmitted to the car using some mobile radio communication like the cellular phone network for example.
Of course, good compression rates are desireable as they help to save costs and transmission time.
Especially on countryside highways bandwidths can be low.

In this paper we describe two approaches for compressing a route:
In the first approach, we represent a route by giving only a few road segments which we call \emph{via edges\nocorr}.
Equivalently, one can also use \emph{via nodes\nocorr}, though this requires some simple preprocessing of the road network.
Via edges and via nodes require that the mobile device located in the car has passably up-to-date map data and the capability to compute optimal routes with respect to this data.
In the second approach, \emph{contraction hierarchies}~\cite{gssd-chfsh-08} are used to encode a route using \emph{shortcuts\nocorr}.
In this case the mobile device in the car has to support contraction hierarchies too.
The two approaches can also be combined.

\section{Preliminaries}

In route planning a road network is usually represented by a weighted directed graph $G=(V,E,c)$ where the nodes represent junctions and the edges represent road segments; $c:E\to\mathcal{C}$ assigns a weight to every edge in $E$.
Here, we only consider the simplest case of time-\emph{in}dependent edge weights, where $\mathcal{C}=\real_{>0}$ holds.
But of course, our method can be applied to more general scenarios, where the server performs time-dependent route planning~\cite{bdsv-tdch-09,bgns-tdcha-10} for example.

For $\mathcal{C}=\real_{>0}$ optimal routes can be computed using Dijkstra's algorithm.
Alternatively, one of the numerous recent and much more efficient methods can be used~\cite{dssw-erpa-09}.
Routes are actually paths in $G$.
Here, we formalize a path as a sequence $\langle (u_1,u_2),(u_2,u_3),\dots,(u_n,u_{n+1}) \rangle$ with $(u_i,u_{i+1})\in E$ for $1\leq i\leq n$.
Of course, two paths $P:=\langle (u_1,u_2),\dots,(u_n,u_{n+1}) \rangle$ and $P':=\langle (v_1,v_2),\dots,(v_m,v_{m+1}) \rangle$ with $u_{n+1}=v_1$ can be concatenated --- we write $P\oplus P'$ for the concatenated path.
Shortest paths are not necessary unique in $G$, i.e., there may be more than one shortest path between the respective start and destination nodes.
But if there is a unique shortest path from a node $u$ to a node $v$, then we denote it by $\SP(u,v)$.

\section{Compression of Paths Using Via Edges}

\subsection{Representing a Path by Via Edges}
Given a path $P=\langle (u_1,u_2),(u_2,u_3),\dots,(u_n,u_{n+1})\rangle$ in $G$, which is not necessarily a shortest path.
Let
$Q=\langle (u_{i_1}, u_{i_1+1}), (u_{i_2}, u_{i_2+1}), \dots, (u_{i_k}, u_{i_k+1})\rangle$
be a subsequence of $P$ with the property that the shortest paths
from $u_1$ to $u_{i_1}$, from $u_{i_k+1}$ to $u_n$,
and from $u_{i_j + 1}$ to $u_{i_{j+1}}$ are unique for all $1\leq j < k$.
If
\[
  P\ =\ 
    \SP(u_1,u_{i_1})\oplus \langle(u_{i_1},u_{i_1 + 1})\rangle \oplus \SP(u_{i_1 + 1}, u_{i_2}) \oplus
    \langle(u_{i_2}, u_{i_2 + 1})\rangle \oplus \dots \oplus
    \SP(u_{i_k + 1},u_{n+1})
\]
holds, then $P$ is completely determined by $Q$.
We call $Q$ a \emph{representation of $P$ by via edges} in this case.
There, $(u_{i_1},u_{i_1 + 1}),\dots,(u_{i_k},u_{i_k + 1})$ are the \emph{via edges\nocorr}.

\begin{mycorollary}
  \label{crl:via-edges-leq}
  If $Q$ is a representation of a path $P$ by via edges, then $|Q|\leq |P|$ holds.
\end{mycorollary}

Corollary~\ref{crl:via-edges-leq} means that $Q$ is a compressed representation of $P$.
At best, $P$ itself is a unique shortest path in $G$ and we can represent $P$ by the empty sequence $Q=\langle\rangle$
if the start and the destination of $P$ are known.
Given a representation $Q=\langle(u_{i_1},u_{i_1 + 1}),\dots,(u_{i_k},u_{i_k + 1})\rangle$ of a path $P$ by via edges the path $P$ can be reconstructed very easily:
We only have to perform a one-to-one shortest path query between all pairs of nodes $u_{i_j + 1}$ and $u_{i_{j+1}}$ for $j\in\{1,\dots,k-1\}$, between the start node and $u_{i_1}$, as well as between $u_{i_k + 1}$ and the destination node.

\subsection{Representing Paths by Via Nodes}
\label{sec:via-nodes}
Analogous to via edges, several paths can also be represented by \emph{via nodes\nocorr}.
Then, a path $P$ is represented as a sequence of \emph{nodes} lying on $P$.
Again, the subpaths between the via nodes have to be unique shortest paths.
However, this fails in some cases.
Consider, for example, a path that consists of one single edge and that is not a unique shortest path.
With via nodes we could not represent this path.
However, this problem can solved if we apply a simple preprocessing to $G$:
We just have to replace all edges in $(u,v)\in E$, that are not unique shortest paths, by a path $\langle(u,w_{\mathrm{new}}),(w_{\mathrm{new}}, v)\rangle$, where $w_{\mathrm{new}}$ has to be newly added to $V$.
Also, we have to set $c((u,w_{\mathrm{new}})):=c((w_{\mathrm{new}},v)):=c((u,v))/2$.
Having applied this preprocessing all shortest paths in $G$ can be represented using via nodes.
Although the results in the work are described in terms of via edges, they can also be used with via nodes ---
if this preprocessing is applied to the underlying graph before.

\subsection{Computing a Minimal Representation by Via Edges}
\label{sec:def-via-edges}
A representation $Q$ of a path $P$ by via edges is \emph{minimal} if for all other such representations $Q'$ holds $|Q|\leq |Q'|$.
Algorithm~\ref{alg:via-edges-linear} yields such a minimal representation of a path $P$ by via edges.
It performs $\Oh(|P|)$ one-to-one shortest path queries in $G$.

\begin{algorithm}[t]
\DontPrintSemicolon
  \Function{viaEdges}{$\langle (u_1,u_2),\dots,(u_n,u_{n+1}) \rangle:\mathit{path}$}{\textit{sequence} \textbf{of} $E$}{
  $i:=1 : \nat$ \;
  $Q:=\langle \rangle : \mathit{sequence}\ \mathbf{of}\ E$ \;
  \For{$j:=1$ \KwTo $n$}{
    \lIf{$\langle (u_i,u_{i+1}),\dots,(u_j,u_{j+1}) \rangle$ is unique shortest path from $u_i$ to $u_{j+1}$ in $G$} \textbf{continue}\;
    $Q.\mathit{append}((u_j,u_{j+1}))$ \;
    $i:=j+1$ \;
  }
  \Return $Q$ \;
}
\caption{
  Computes a minimal representation of a path by via edges.
}
\label{alg:via-edges-linear}
\end{algorithm}

Algorithm~\ref{alg:via-edges-log} does in principle the same as Algorithm~\ref{alg:via-edges-linear}
because both algorithms compute the maximum $p\in\{i-1,\dots,n\}$ such that the subpath $\langle (u_i,u_{i+1}),\dots,(u_p,u_{p+1}) \rangle$ of $P$ is a unique shortest path in $G$ (there, $p=i-1$ means that no such subpath of $P$ starting at $u_i$ exists).
Then, both algorithms append the edge $(u_{p+1},u_{p+2})$ to the result sequence $Q$.
In case of Algorithm~\ref{alg:via-edges-log} we delegate the computation of $p$ to the auxiliary function \textit{maxPrefixSP} (cf. Lines~\ref{lin:prefix-begin} to~\ref{lin:prefix-end}) which works similar to binary search and performs only $\Oh(\log|P|)$ one-to-one shortest path queries hence.
As a consequence Algorithm~\ref{alg:via-edges-log} computes the same result as Algorithm~\ref{alg:via-edges-linear} but
needs only $\Oh(|Q|\log|P|)$ one-to-one shortest path queries for $Q$ being the resulting representation by via edges.
This means that Algorithm~\ref{alg:via-edges-log} should run faster than Algorithm~\ref{alg:via-edges-linear} when $Q$ contains only few edges.
If we replace the function \textit{maxPrefixSP} from Algorithm~\ref{alg:via-edges-log} by another variant (cf. Algorithm~\ref{alg:faster-prefix}), we can even reduce the number of one-to-one shortest path queries to $\Oh(|Q|\log(|P|/|Q|))$.

\begin{algorithm}[t]
\DontPrintSemicolon
\Function{maxPrefixSP\nllabel{lin:prefix-begin}}{$j,k:\nat$}{$\nat$}{
  $(\ell,r):=(j-1,k)$\;
  \While{$\mathit{true}$}{
    \lIf{$\ell + 1 = r$} \Return $\ell$\;
    $m:=\lfloor (\ell+r)/2 \rfloor$\;
    \lIf{$\langle (u_j,u_{j+1}),\dots,(u_m,u_{m+1}) \rangle$ is unique shortest path from $u_j$ to $u_{m+1}$ in $G$} $\ell:=m$\;
    \lElse $r:=m$\;\nllabel{lin:prefix-end}
  }
}
\BlankLine
\Function{viaEdges}{$\langle (u_1,u_2),\dots,(u_n,u_{n+1}) \rangle:\mathit{path}$}{\textit{sequence} \textbf{of} $E$}{
  $i:=1 :\nat$\;
  $Q:=\langle\rangle : \mathit{sequence}\ \mathbf{of}\ E$\;
  \While{$i\leq n$}{
    $p := \textit{maxPrefixSP}(i,n)$\;
    \lIf{$p<n$} $Q.\mathit{append}((u_{p+1},u_{p+2}))$\;
    $i:=p+2$\;
  }
  \Return $Q$\;
}
\caption{
  Computes the same as Algorithm~\ref{alg:via-edges-linear} but runs faster when few via edges are needed.
}
\label{alg:via-edges-log}
\end{algorithm}

\begin{algorithm}[t]
\DontPrintSemicolon
\Function{maxPrefixSP}{$j,k:\nat$}{$\nat$}{
  $h:=0$\;
  \While{$\mathit{true}$}{
    \lIf{$\langle (u_j,u_{j+1}),\dots,(u_{j+h-1},u_{j+h}) \rangle$ is unique shortest path from $u_j$ to $u_{j+h}$ in $G$}
      $h:=2h$\;
    \lElse{\textbf{break}}
  }
  $(\ell,r):=(j+\lfloor h/2\rfloor-1,k)$\;
  \While{$\mathit{true}$}{
    \lIf{$\ell + 1 = r$} \Return $\ell$\;
    $m:=\lfloor (\ell+r)/2 \rfloor$\;
    \lIf{$\langle (u_j,u_{j+1}),\dots,(u_m,u_{m+1}) \rangle$ is unique shortest path from $u_j$ to $u_{m+1}$ in $G$} $\ell:=m$\;
    \lElse $r:=m$\;
  }
}
\caption{Reduces the asymptotic running time of Algorithm~\ref{alg:via-edges-log} to $\Oh(|Q|\log(|P|/|Q|))$.}
\label{alg:faster-prefix}
\end{algorithm}

\section{Applying Contraction Hierarchies}
\label{sec:compress-ch}

A contraction hierarchy (CH)~\cite{gssd-chfsh-08,g-ch-08} is a hierarchical representation of a road network, where the nodes are ordered by \emph{importance\nocorr}.
More important nodes are on a higher level of the CH than less important nodes.
The CH is obtained from the original road network $G$ in a preprocessing step.
It contains \emph{shortcut edges} that leap valleys, i.e., subpaths that go through lower levels.
More precisely, every shortcut edge represents a path of the form $\langle (v,u),(u,w) \rangle$ such that the level of $u$ is lower than the level of $v$ and $w$.
Note, that $(v,u)$ and $(u,w)$ can be shortcuts themselves.

\subsection{Compressing a Path Using Contraction Hierarchies}

Originally, CHs are used for the very fast computation of shortest paths.
Here, we also use them to represent paths, that are not necessary shortest ones, in a space efficient way:
Given a path $P=\langle (u_1,u_2),(u_2,u_3),\dots,(u_n, u_{n+1}) \rangle$ and a CH we compress $P$ by replacing subpaths of $P$ with shortcuts present in the CH.
To do so, we ``contract'' the path $P$ as far as possible.
Algorithm~\ref{alg:compress-using-CH} shows this method in more detail.
Note, that even if a shortcut $(v,w)$ is present in the CH, it is not necessarily the case that this shortcut represents the path $\langle (v, u_{i_j}), (u_{i_j},w) \rangle$.
This is because parallel edges arise during the construction of the CH and parallel edges are thrown away during preprocessing.

Applying Algorithm~\ref{alg:compress-using-CH} to a path $P$ we get a path $P'$ which contains shortcuts.
If we repeatedly replace all shortcuts in $P'$ by the path they represent until no shortcuts are left, then we surely end up with the original path $P$.
Hence, $P$ is completely determined by $P'$ and we call $P'$ a \emph{representation of $P$ using a CH\nocorr}.
Note, that $P'$ as generated by Algorithm~\ref{alg:compress-using-CH} is minimal with respect to this property.
More precisely, for every other path $P''$, that contains shortcuts from the given CH and that completely determines $P$, we have $|P'|\leq |P''|$.
According to the following corollary such a representation is also a way to compress a path:

\begin{mycorollary}
Let $P$ be a path in $G$.
Let $P'$ be the result of Algorithm~\ref{alg:compress-using-CH} when applied to $P$ with some CH given.
Then $|P'|\leq |P|$ holds.
\end{mycorollary} 

\begin{algorithm}[t]
\DontPrintSemicolon
\Function{compressWithCH}{$\langle (u_1,u_2),\dots,(u_n,u_{n+1})\rangle:\mathit{path}$}{\textit{path}}{
  $R:=\langle (u_1,u_2),\dots,(u_n,u_{n+1})\rangle : \mathit{path}$\;
  sort $u_2,\dots,u_n$ according to increasing level yielding $u_{i_1},\dots,u_{i_{n-1}}$\nllabel{lin:sort-nodes-of-path}\;
  \For{$j:=1$ \KwTo $n-1$\nllabel{lin:for-loop-start}}{
    \If{a shortcut $(v,w)$ representing the subpath $\langle (v, u_{i_j}), (u_{i_j}, w)\rangle$ of $R$ is present in the CH}{
      replace $\langle (v, u_{i_j}), (u_{i_j}, w)\rangle$ by $\langle (v,w) \rangle$ in $R$\nllabel{lin:for-loop-end}\;
    }
  }
  \Return $R$\;
}
\caption{Replaces subpaths of a path by shortcuts of a CH.}
\label{alg:compress-using-CH}
\end{algorithm}

The sorting of the nodes (cf. Line~\ref{lin:sort-nodes-of-path}) can be done in $\Oh(|P|\log |P|)$ time.
Let $d$ be the maximum number of incoming or outgoing edges of a node in the CH.
Then, the Lines~\ref{lin:for-loop-start} to~\ref{lin:for-loop-end} take $\Oh(d|P|)$ time.
That means, Algorithm~\ref{alg:via-edges-log} needs $\Oh(d|P|+|P|\log |P|)$ time in total.

\subsection{Combining Via Edges and Contraction Hierarchies}

The representation by via edges and the representation using a CH can be combined quite easily.
First, apply Algorithm~\ref{alg:compress-using-CH} to the given path $P$ and then compute a representation $Q'$ of the resulting path $P'$ by via edges.
The unpacking of $Q'$ is straightforward then:
Just reconstruct $P'$ from $Q'$ as described in Section~\ref{sec:def-via-edges} and then recursively unpack all shortcuts in $P'$ completely.

Note, that $|P'|\leq |P|$ holds.
We even expect that in many cases we have $|P'|\ll |P|$.
But then, we have to perform much less one-to-one shortest path queries than if we represent $P$ by via edges directly.
But one-to-one shortest path queries are one of the expensive operations in this setup.
So, the combination of both methods may not only provide better compression but also better running times than if via edges were used alone.

\section{Application}

\subsection{A Hybrid Scenario}
We propose a hybrid scenario of server based mobile routing, where start, destination, and maybe some additional requirements are send from the car over an available wireless communication network to a server.
The message containing this query will be very small.
Unfortunately, this will most probably not be the case for the resulting route, which may consist of hundreds of road segments and must be send from the server to the car.

If routes consist mainly of subpaths which are unique shortest paths, then the sizes of the corresponding messages can be reduced greatly if we represent the routes by via edges. Further compression is possible by using via nodes
which can be represented somewhat more compactly than edges but require the
preprocessing explained in Section~\ref{sec:via-nodes}.

The route has to be reconstructed in the car.
This requires a mobile device present in the car, which is equipped with map data which is not too far away from the current state of the road network and which is able to compute optimal routes with respect to this data.
However, the map data must be present in the car anyway as driving directions have to be generated in the car in this setting.
Of course, the server has to know the map data present in the car exactly.
To achieve this, the server could store every version of the map data which has ever been delivered to any cars mobile device.
The mobile device has to send the unique ID of its map data to the server together with every query.
Note, that outdated map data requires that additional information is send from the server to the car if necessary.
If this happens only rarely, this is no problem.
However, updates of the cars map data may be necessary from time to time.

The compression of routes using via edges or via nodes only makes sense, when the compressed representation needs significantly less space than the route itself.
In this case the mobile device has to perform only few one-to-one shortest path computations to uncompress the route.
However, with Dijkstra's algorithm this would still be to slow.
This holds even more for mobile devices, which usually have little main memory and slow flash storage.
However, the current speedup techniques for shortest-paths computation~\cite{dssw-erpa-09} can solve this problem.
Especially, we recommend contraction hierarchies~(CHs)~\cite{gssd-chfsh-08,g-ch-08}, which were used to compress routes in Section~\ref{sec:compress-ch}.
CHs are a very effective speedup technique, that has already successfully been adapted to the needs of mobile devices~\cite{ssv-mrp-08,v-femno-10}.
Mobile contraction hierarchies need on average 0.1\,sec to answer a random shortest-path query on a data set of the European road network~\cite{ssv-mrp-08}.
So, if a route is represented by 10 via edges, for example, then we have to perform 11 one-to-one shortest path queries and the whole route can be constructed within 1.1\,sec --- or even less as the paths between the via edges are most likely not that long.
Furthermore, to start driving, we only need the \emph{first} part of the route quickly which requires only one shortest-path query.

Note, that on server systems CHs run much faster than on mobile divices:
For the European road network an average one-to-one shortest path query need less than 1\,msec.
As an example consider a route, that consists of 700 edges and that can
be represented by 10 via edges.\footnote{Note, for comparison,
that an average route in the European road network has about 1\,400 edges.
But most real-live routes should be shorter.}
Then, Algorithm~\ref{alg:via-edges-log} performs up to $11\cdot\log 690\approx 104$ one-to-one shortest path queries at less than 1\,msec each.
That means, that the server needs less than 0.1\,sec for compressing the complete route in this example.

As an alternative to via edges we can compress the route using a CH.
In this case, the server needs not only to know the map data present in the car but also the exact structure of the CH present in the car.
Also, we can combine both techniques as described in Section~\ref{sec:compress-ch}.
However, which of the two methods works better and whether the combination brings a significant better compression rate within less running time must be evaluated experimentally.

\subsection{The Benefit of Server-Based Mobile Route Planning}
Server-Based mobile route planning enables several interesting applications.
If the server has detailed statistical data about the usual traffic volume of many streets for every day of the week and for every time of day, then the server can compute a \emph{time-dependent} optimal route~\cite{bdsv-tdch-09,bgns-tdcha-10}.
Or, the server could compute a route which is optimal depending on a certain parameter which can be chosen freely by the driver of the car (we call this \emph{flexible} route planning~\cite{gks-rpfof-10}).
An example of such a parameter is the tradeoff between travel time and fuel consumption.
It is not at all trivial to make time-dependent and flexible route planning run efficiently on a mobile device without communicating with a server.
Also, the server could support \emph{ride sharing}~\cite{glsnv-fdcrs-10}.
Alternative routes are also interesting~\cite{adgw-arrn-10,dgsb-dcarr-10}.

\section{Conclusions and Future Work}
We have presented two methods to compress the description of routes, that is of paths.
The first method represents a path, which is a sequence of edges, by a subsequence of \emph{via edges} such that the missing parts are unique shortest paths.
Alternatively, \emph{via nodes} can be used instead of via edges.
Then, the path requires less space but the road network requires more space.
The second method replaces subpaths of the original path by shortcuts that are taken from a contraction hierarchy.
In contrast to the first method, the second method performs no one-to-one shortest path queries.
As this is a relatively expensive operation, we expect that compression and uncompression take less time for the second method.
However, the two methods can also be combined.
This way we hope to achieve better compression within less running time than if first method were used alone.

We also described a possible scenario:
The routes are computed on a server, which enables us to take the latest data about congestions and changes of the road network into account.
Then we have to transmit a description of the route to the car using some mobile communication network.
As a route can consists of hundreds of edges, the accordant message can get quite large.
This is where we apply the compression methods presented in this work.
We hope to reduce the size of the message significantly ---
not only to save costs but also to make our approach applicable even when bandwidths are low.
We have not evaluated these ideas with realistic data yet, but we hope to do this in the not too far future.

If the routes are computed on a server rather than in the car, many more interesting applications are possible.
These are time-dependent routeplanning~\cite{bdsv-tdch-09,bgns-tdcha-10}, flexible route planning~\cite{gks-rpfof-10}, support for ride-sharing~\cite{glsnv-fdcrs-10}, and alternative routes~\cite{adgw-arrn-10,dgsb-dcarr-10} for example.


\bibliographystyle{plain}
\small
\bibliography{references}

\end{document}